# Application of Novel PACS-based Informatics Platform to Identify Imaging Based Predictors of CDKN2A Allelic Status in Glioblastomas

Niklas Tillmanns[1,4], Jan Lost[1], Joanna Tabor[2], Sagar Vasandani[2], Shaurey Vetsa[2], Neelan Marianayagam[2], Kanat Yalcin[2], E. Zeynep Erson-Omay[2], Marc von Reppert[1], Leon Jekel[1], Sara Merkaj[1], Divya Ramakrishnan[1], Arman Avesta[3], Irene Dixe de Oliveira Santo[1], Lan Jin[4], Anita Huttner[5], Khaled Bousabarah[6], Ichiro Ikuta[7], MingDe Lin[1,8], Sanjay Aneja[3], Bernd Turowski[9], Mariam Aboian[1*], Jennifer Moliterno[2*]

[1] Brain Tumor Research Group, Department of Radiology and Biomedical Imaging, Yale School of Medicine, 333 Cedar Street, PO Box 208042, New Haven, CT 06520, USA

[2] Department of Neurosurgery, Yale School of Medicine, New Haven, Connecticut, USA.

[3] Department of Radiation Oncology, Yale School of Medicine, 333 Cedar Street, PO Box 208042, New Haven, CT 06520, USA

[4] R&D, Sema4, 333 Ludlow Street, North Tower, 8th floor, Stamford, CT 06902, USA

[5] Department of Pathology, Yale School of Medicine, New Haven, Connecticut, USA

[6] Visage Imaging, GmbH., Lepsiusstraße 70, 12163, Berlin, Germany

[7] Mayo Clinic Arizona, Department of Radiology, 5711 E Mayo Blvd, Phoenix, AZ 85054, USA

[8] Visage Imaging, Inc., 12625 High Bluff Dr, Suite 205, San Diego, CA 92130, USA

[9] University Dusseldorf, Medical Faculty, Department of Diagnostic and Interventional Radiology, D-40225 Dusseldorf, Germany

*Corresponding author:*

Mariam S. Aboian, M.D./Ph.D.,

(650) 285-7577, mariam.aboian@yale.edu,

789 Howard Avenue (CB30), PO Box 208042, New Haven, CT 06520, United States of America

**Abstract:**

Gliomas with CDKN2A mutations are known to have worse prognosis but imaging features of these gliomas are unknown. Our goal is to identify CDKN2A specific qualitative imaging biomarkers in glioblastomas using a new informatics workflow that enables rapid analysis of qualitative imaging features with Visually AcceSAble Rembrandtr Images (VASARI) for large datasets in PACS.

Sixty nine patients undergoing GBM resection with CDKN2A status determined by whole-exome sequencing were included. GBMs on magnetic resonance images were automatically 3D segmented using deep learning algorithms incorporated within PACS. VASARI features were assessed using FHIR forms integrated within PACS.

GBMs without CDKN2A alterations were significantly larger (64% vs. 30%, p=0.007) compared to tumors with homozygous deletion (HOMDEL) and heterozygous loss (HETLOSS). Lesions larger than 8 cm were four times more likely to have no CDKN2A alteration (OR: 4.3; 95% CI:1.5-12.1; p<0.001).

We developed a novel integrated PACS informatics platform for the assessment of GBM molecular subtypes and show that tumors with HOMDEL are more likely to have radiographic evidence of pial invasion and less likely to have deep white matter invasion or subependymal invasion. These imaging features may allow noninvasive identification of CDKN2A allele status.

Keywords:

Glioblastoma, Machine learning, FHIR, PACS-integration

# **Introduction:**

Glioblastoma (GBM) is the most common primary brain tumor in adults and accounts for 15% of all brain tumors. It occurs with an incidence of 3.22 per 100.000 cases in the United States annually.[1] The current standard of care treatment for glioblastoma (GBM, IDH-wild type, WHO Grade 4) is maximum surgical resection, followed by chemo- and radiotherapy.[2] Gliomas are classified according to the WHO classification of central nervous system tumors with a recently published version in 2021[3] differentiating adult-type diffuse gliomas into three entities: astrocytoma (IDH mutant, 1p19q intact), oligodendroglioma (IDH mutant, 1p19q codeletion), and glioblastoma (IDH wildtype).[3] This new classification diagnoses gliomas not solely based on histology, but complemented by more sophisticated molecular markers such as *CDKN2A* which is important for *IDH*-mutant gliomas.

Patients with *IDH*-mutant gliomas that present with either homozygous or heterozygous *CDKN2A* deletion have decreased progression-free and overall survival.[4,5]

Cyclin-dependent kinase inhibitor (*CDKN*) is a gene located on chromosome 9p21.[4,6,7] *CDKN* has two subtypes, *CDKN2A* and *CDKN2B* (locus (INK4a/ARF), that encode for tumor suppressor proteins *(p14$^{ARF}$ and p16$^{INK4A}$*) inhibiting the transition from G1-phase to S-phase in the cell cycle. *p14$^{ARF}$* activates *p53,* which results in the inhibition of cell growth.[4,6,7]

Conversely, CDKN2A homozygous deletion in GBM has a less established role, and is not included in the WHO 2021 criteria. However, studies support the hypothesis of CDKN2A homozygous deletion determining a worse prognosis in GBM[8], and suggest that GBM with CDKN2A homozygous deletion may benefit from higher dose radiation[8]. This presents a critical need for predicting this molecular subtype of glioblastomas.

In order to make these urgent information available as soon as possible, development of standardized imaging biomarkers is necessary. In comparison to biopsy, MRI is a routine and noninvasive procedure, in which it can not only help decreasing the risk of biopsies.[9,10] But also in helping to establish a diagnosis in tumors that are not feasible to biopsy. In addition, it can help clinicians make treatment decisions given the heterogeneity of the tumor and the known limitations of biopsies in this regard by evaluating the whole tumor.[11,12] As a single biopsies may lead to underestimation of the genetic variance in the tumor and therefore to an incomplete therapy.

Standard of care pre-operative imaging of glioblastomas on MRI includes multiple sequences: T2, FLAIR, and T1 with and without gadolinium-based contrast agent sequences.[13] Therefore, the determination of biomarkers from widely-used imaging sequences will be most applicable to routine clinical practices and circumvent the lack of widespread availability of advanced imaging modalities (such as tumor perfusion-weighted imaging).

To determine meaningful and reliable MRI features, a comprehensive and standardized feature set is needed to ensure reproducibility. Performing imaging phenotype analysis of brain tumors can be very time consuming and requires handling of multiple software packages, which limits the ability to evaluate phenotypes in rapidly available timeframe. Development of informatics tools that allow phenotype assessment within the same platform, can dramatically expedite the phenotypic classification and allow generation of valuable descriptive information in the rapidly progressing field of brain tumor classification. We used VASARI features to determine qualitative imaging features unique for GBMs with *CDKN2A* alterations, which are currently not well understood and may be of interest to be subclassified in future cIMPACT guidelines or WHO criteria.[8,14] VASARI stands for Visually AcceSAble Rembrandt Images, and is a comprehensive MRI feature set scheme for reproducible measurement of brain tumors.[15] The feature set consists of 29 scoring items with a defined lexicon to ensure a standardized and consistent assessment of non-contrast and contrast-enhanced MR images (Supplementary Data 1).[16] VASARI was developed by a working group of multiple neuroradiologists from different institutions to ensure maximal applicability to brain tumor imaging, and made freely available by several radiological organizations.[17] The features were validated in a consensus group of 8 radiologists.[15] In recent works, proportional VASARI features including the percentage of total abnormal tissue classified as contrast- enhanced tumor, nonenhanced tumor, necrosis, and edema were shown to predict IDH mutation status in GBM preoperatively and served as the reference standard for comparing visual assessment of volume to manually or automatically segmented volumes.[15,16,18,19] VASARI was also used for reproducible molecular profiling in IDH, 1p19q, and EGFR from pre-operative MRI, as well as predicting predict molecular profiles in glioblastoma based on VASARI. [15,20].

To our knowledge, VASARI has not yet been successfully used to assess *CDKN2A* homozygous deletion (HOMDEL) status in GBM according to WHO 2021. The practical implementation of VASARI is laborious therefore we evaluated the feasibility of clinical incorporation of VASARI forms in a streamlined workflow using Fast Health Interoperability Resources (FHIR) forms. FHIR is a medical information processing and communication standard that works on a questionnaire and response system and provides easy a user friendly interface through the NIH website.[21] Incorporation of FHIR into PACS allows direct linking electronic medical data and qualitative data analysis with DICOM format of images. This incorporation of informatics tools into one software package is the basis for a relational database approach for brain tumor analysis and was critical for our phenotypic characterization of glioblastomas based on *CDKN2A* HOMDEL status.

# **Objective:**

We aim to identify qualitative imaging biomarkers specific for *CDKN2A* deletion in GBMs using a novel informatics workflow that allows fast analysis of qualitative imaging features using VASARI for large datasets from an integrated database that incorporates DICOM images with FHIR format information.

**Methods**

The dataset contains 69 newly diagnosed patients from our institution. All patients underwent primary surgery in 2021 for glioblastoma characterized by WHO 2021 criteria and consented for whole exome sequencing to be performed on available tissue. The study was approved by the Yale University IRB and need for consent was waived. All methods were carried out in accordance with relevant guidelines and regulations.

IRB waiver of informed consent was obtained for all patients who underwent resection for glioblastoma from January 2020 to December 2021 at Yale-New Haven Health and retrospectively reviewed. We included all patients with known *CDKN2A* deletion status, determined by whole exome sequencing, and grouped these according to the number of *CDKN2A* copies. Further inclusion criteria were the availability of pre-operative MRI with either FLAIR + T1 post-gadolinium spin echo (PGSE) or FLAIR + T1 post-gadolinium gradient echo (PGGE) sequences. IDH-mutant gliomas were excluded.

Magnetic resonance images were transferred from the clinical PACS to the research PACS (AI Accelerator, Visage Imaging, Inc. San Diego, CA). Deep learning-based automatic segmentation built within PACS was used for tumor segmentation.[22] Specifically, a UNETR deep learning algorithm used FLAIR and T1 post gadolinium sequences to segment the Whole, Core, and Necrotic portions of the tumor according to BraTS criteria.[23] For further information on the algorithm pipeline we refer to one of our prior publication.[22] Two medical student research fellows (NT, JL) revised the segmentations, which were then validated and revised as needed by a board-certified neuroradiologist (MSA). Extracted features included: percent edema, percent contrast enhancement, and percent necrosis, which were calculated based on volumetrics described above and reported into the respective VASARI categories. As done in prior studies, visual-based estimations of these percentages were not performed due to the known potential for the inaccuracy of the results.[15] The VASARI form was scored by a board-certified neuroradiologist (MSA) in PACS through a custom built-in Health Level 7® Fast Healthcare Interoperability Resource® (HL7 FHIR) webform. The workflow was streamlined as the neuroradiologist opened a study in our research PACS. Within the interface, there is a button called "VASARI" which can be clicked. This opens the FHIR form with the VASARI questionnaire in it. The FHIR form is opened right next to the PACS viewer in a separate window through a link within PACS, the MRI study can be scrolled and the VASARI questions can be answered. The VASARI feature set consists of 29 scoring items with a defined lexicon to ensure a standardized and consistent assessment.[16] Checkbox fields were used as input field type for VASARI scoring. At the end of the questionnaire there was a freeform text field for additional information (Supplementary Fig. 1). After all patients were scored, the completed fields within the FHIR forms were then exported into Excel (Microsoft, Redmond, WA) for statistical analysis.

Fifteen cases that were randomly selected were evaluated to compare the efficiency for scoring VASARI features from opening the study to the completion of scoring using the traditional manual assessment and data entry vs. automatic assisted assessment and FHIR form data entry. For the automated analysis, the studies were opened in PACS and scored by a board-certified neuroradiologist (MSA) while the time and clicks per case were assessed manually (NT,JL). For the manual scoring, the studies were opened in PACS and manually scored in a separate Excel document by a board-certified neuroradiologist (MSA) while the time and clicks per case were assessed manually (NT,JL). The evaluation included the number of clicks per case and time per case.

Statistical analysis:

Descriptive statistics of radiogenomic features were summarized by the 3 subgroups of CDKN2A. Based on the distributions of these features, we classified CDKN2A subgroups, and conducted statistical testing to investigate the differences in the features between the reclassified subgroups. For the correlations between subgroups and features, Fisher's Exact Test was used for categorical variables, while Student's t test or Mann-Whitney U test was used for continuous variables based on the distribution (Supplementary Table 2). For features that proved to be statistically significant in a first univariate analysis and showed to discriminate certain subgroup from others, we developed logistic regression models to predict the relevant subgroup.

Genomic analysis:

To detect somatic single-nucleotide variations (SNVs), insertions/deletions (INDELs), and Copy Number Aberrations (CNAs), Whole Exome Sequencing was performed on the tumor samples acquired from the OR along with their matching blood samples to be used as normal. Sequencing was performed at the Yale Center for Genome Analysis using the Illumina NovaSeq 6000 system with $2 \times 101$–bp reads following the capture of the regions using IDT xGen, IDT GOAL or Roche_MedExome panels. Average mean coverages of $109.2\times$ and $214.0\times$ were achieved for blood and tumor tissues, respectively. Somatic variant calling for SNVs/INDELs along with variant annotation was performed as previously described in reports from our institution[24]. Copy number aberrations were determined using an in-house script using the ratio of tumor/normal coverage, normalized by total coverage variation and segmentation, performed using DNAcopy R package[25]. Copy-neutral loss of heterozygosity (LOH) was determined by using the deviation of Variant Allele Frequency (VAF) for germline heterozygous mutations in tumor compared to blood.

# Results

Patient characteristics:

Among the 69 patients included in the final analysis, there were 25 tumors (36%) that had *CDKN2A* heterozygous deletion (HETLOSS), 17 tumors (25%) had biallelic loss (HOMDEL), and the remainder, 37 (39%) had intact copy numbers (Figure 1). The cohort contained 44 males (64%) and 25 females (36%). 25 patients had heterozygous loss of CDKN2A (36%). 17 patients had homozygous deletion (25%), while the rest presented with no alteration of CDKN2A status (27, 39%). EGFR amplification was found in 42 (61%) patients (Table 1).

Qualitative imaging features analysis:

Patients with HOMDEL of *CDKN2A* exhibited lower levels of deep white matter invasion (47.1%), defined as "Enhancing or nCET tumor extending into the internal capsule, corpus callosum or brainstem" compared to those with HETLOSS or no alteration (75%) (p= 0.041). HETLOSS and no alteration groups also had higher subependymal invasion (87% vs. 59%, p=0.032) defined as "Invasion of any adjacent ependymal surface in continuity with enhancing or non-enhancing tumor matrix" than HOMDEL.

A lower percentage of pial invasion was found in the HETLOSS, and no alteration groups (52% vs. 82%, p=0.045) compared to HOMDEL. The pial invasion was predictive of HOMDEL (OR: 8.1, 95% CI: 1.8-53.2; p<0.012) as tumors with pial invasion were eight times more likely to be HOMDEL, even after adjusting for deep white matter and subependymal invasion as covariables in the logistic regression model (Fig 2 & Fig 3). The model did not improve by incorporating other qualitative imaging features in the analysis.

GBMs without *CDKN2A* alterations were significantly larger in size when compared to tumors with HOMDEL and HETLOSS (64% vs. 30%, p=0.007). The direct comparison of whole tumor volume that includes a non-enhancing portion of the tumor defined by FLAIR among the wildtype, HOMDEL, and HETLOSS is shown in Figure 2. Lesions greater than 8 cm were four times more likely to be found in patients without alteration of *CDKN2A* (OR: 4.3; 95% CI:1.5-12.1; p <0.001) compared to HOMDEL or HETLOSS. 8 cm were defined as the largest (x-y) cross-sectional diameter of T2 signal abnormality measured on a single axial image according to VASARI.

Manual VASARI scoring vs. built-in FHIR form:

Fifteen cases were evaluated to compare the time for scoring VASARI features from opening the study to completion of scoring. The time for automated measurements was 2.76 min (SD ± 0.47), and for manual measurements, 5.91 min (SD ± 0.87). The difference between automatic and manual measurements was statistically significant (p<0.0001) using an unpaired t-test (Fig. 4). This highlights the workflow inefficiencies of manual assessment of VASARI forms using separate scoring modalities compared to native, built-in FHIR (Fast Healthcare Interoperability Resources) forms within PACS. This is supported by the amount of clicks needed per case from opening the study to completion of scoring between built in analysis and the manual group. The mean amount of clicks for automated measurements was 43.80 (SD ± 6.268), and for manual measurements, 76 (SD ± 6.245). The difference between built-in and manual measurements was statistically significant (p<0.0001) using an unpaired t-test (Fig. 4).

## Discussion:

The 2021 WHO classification identified novel molecular subtypes, including *CDKN2A* homozygous deletion status in gliomas. But recent literature suggests that *CDKN2A* homozygous deletion status can also predict worse outcomes in patients with GBM which are IDH-wildtype.[2,5,8,14,26,27]

These findings are not yet incorporated in clinical patient care, since most patient with WHO grade 4 tumors are treated with the same therapy. Nonetheless early identification of CDKN2A status might lead to a more aggressive approach in surgery or higher dose radiotherapy[8] and might allow for inclusion in clinical trials. It will be even more valuable by the time targeted therapies for this specific subtype are incorporated in patient management.[27] Because of the shift towards molecular profiling in glioma diagnosis, and the integration of molecular subtypes in the most recent WHO criteria gathering these information is critical.

We aim to establish correlation of radiological findings and specific genetic alterations to support further clinical decision making. Our study investigated whether qualitative imaging biomarkers for *CDKN2A* can be identified in glioblastomas on pre-operative MR images using standard imaging protocol, as this sub-classification of glioblastomas is currently not available. To our knowledge, this is the first study to attempt to identify such imaging biomarkers in a cohort of glioblastomas with *CDKN2A* alterations.

We found GBMs with homozygous *CDKN2A* loss are more likely to exhibit radiographic evidence of pial invasion and less likely to have deep white matter or subependymal invasion. In addition, tumor volume is also predictive, with tumors greater than 8 cm being less likely to harbor an underlying *CDKN2A* copy loss. These imaging characteristics serve as a non-invasive pre-operative method to measure *CDKN2A* allelic status.

Our findings corroborate with other studies which showed that the prediction of IDH and 1p/19q mutation based on lesion size VASARI features can yield an AUC of 0.73 ± 0.02 and 0.78 ± 0.01, respectively.[19] While these results are promising, the lack of a large volume of literature on this method could be due to the time-intensive nature of performing VASARI scoring. To improve the workflow of VASARI scoring, we leveraged a novel informatics approach using FHIR within PACS to input data more efficiently and quickly into a relational database. Our method includes incorporating ML algorithms into the research version of our clinical PACS, which allows auto-segmentation of tumors using a deep learning algorithm (UNETR). This quantitative method provides higher accuracy of volumetric assessment than the standard VASARI assessment based on qualitative estimation of tumor percent edema, contrast enhancement, and necrosis. As described in prior research, scoring of VASARI is a robust assessment for qualitative assessment of imaging features in gliomas and shows little interobserver variability.[15,19,28] Our PACS embedded software creates an important time and workflow efficacy gain for clinicians and researchers.[29] VASARI integration within PACS provides a streamlined approach for qualitative image assessment that can be integrated into clinical practice."

To date, two-dimensional measurements have been used in routine clinical practice. However, the RANO group has proposed two-dimensional and volumetric measurement protocols for clinical trials.[13] In our study, we performed the most comprehensive evaluation of glioblastoma by including both two-dimensional and volumetric measurements. Nevertheless, volumetric tumor size alone is not sufficient to predict CDKN2A mutation status, as shown by the largely overlapping boxplots (Fig. 2B) and investigation of more complex imaging features like radiomics might be of interest in further studies.[30]. Nonetheless, the results shown above can provide guidance to clinicians so that they are not misled by tumor size, since CDKN2A-mutated tumors with associated poor survival prognosis are often smaller than CDKN2A-intact tumors.[8,14]

Our study highlights the benefits of incorporation of advanced informatics tools to create the relational datasets linked to DICOM images using FHIR standards. FHIR is an emerging and rapidly evolving medical information processing and communication standard, which works on a questionnaire and response system. It can easily exchange and standardize protected health information (PHI) in EMR systems such as EPIC. It is based on Health Level 7 (HL7), a framework of standards for electronic health information exchange, and works with different standardized categories called "Resources."[31] FHIR uses standardized semantics and thus can be easily queried, unifying the way personal health information (PHI) gets acquired and exchanged between different instances in the medical sector. Up to now, incorporating FHIR with DICOM images has not been done, and FHIR is predominantly used in non-imaging workflows. FHIR is expected to be the emerging standard in the coming years to make medical information more accessible for AI applications in the medical sector.[32]

In our approach, we implemented the VASARI scoring through an embedded FHIR form in PACS and were able to decrease the amount of time and clicks per case significantly. Usually, VASARI scoring is done in multiple applications. The radiologist needs to open the study on the PACS station and score VASARI in a separate application like Excel. This not only takes more time and more switching between applications but also hinders the natural workflow and is susceptible to typographical errors. As a result, we created a relational database by the implementation of FHIR forms, which links the patient imaging to the related imaging features and thereby allows for easy organization of larger datasets and the ability to data mine.

Our approach of combining novel informatics methods to build relational databases, machine learning auto-segmentation tools within clinical PACS, and advanced genomic analysis of glioblastomas for a novel biomarker of tumor aggressiveness is a significant advance for the field of neuro-oncology.[29,33] These methods allow the generation of large datasets of annotated images with metadata information on patient outcomes, genetic testing, pathologic results, and detailed qualitative imaging analysis in a streamlined workflow. This workflow has the potential for rapid evaluation of image biomarkers that correlate to several different genetic variants within intracranial malignancies and will overcome the current limitation of extensive human hours required to do this research outside of this workflow. This workflow can also serve as a new and accurate standard for volumetric assessments and will decrease the effort for time-intensive response assessments like RANO and RECIST in routine clinical practice and clinical trials.

Limitations of the study are the small sample size of *CDKN2A* tested GBMs, although this is the largest study assessing standardized imaging features of this molecular subtype in glioblastomas to date.[34] Also the possibility of EGFR status or MGMT status influencing pial -, white matter - or subependymal invasion limits the results, even though it showed to be no significant confounder in our analysis. Future studies with larger sample sizes are needed. The clinician did not score their experience using a questionnaire using a standardized method. Future study investigating the physician perception of using FHIR forms for image annotation is needed. Limited availability of whole exome sequencing results is one of the major contributors to the lack of literature on this topic. Our integrated approach to genomic assessment and imaging correlation represents a strength that allowed the generation of this dataset.

In conclusion, we use a multimodal and multidisciplinary collaborative approach to combine advanced genetic analysis of GBMs and correlate it with image-based analysis accelerated by informatics and machine learning tools to identify imaging biomarkers for *CDKN2A* co-deletion. These imaging biomarkers include tumor size greater than 8 cm and evidence of pial invasion.

**Author contributions:**

All authors contributed to the study conception and design. Material preparation, data collection and analysis were performed by N.T., J.L. and S.V.. The first draft of the manuscript was written by N.T. and all authors commented on previous versions of the manuscript. All authors reviewed and approved the final manuscript.

**Funding:**

The authors declare that no funds, grants, or other support were received during the preparation of this manuscript.

**Competing Interests:**

ML is an employee and stockholder of Visage Imaging, Inc. KB is an employee of Visage Imaging, GmbH. The other authors declare no conflict of interest.

**Data availability statement:**

The datasets analysed during the current study are not publicly available yet, but will be made available in the near future. Currently the dataset is available from the corresponding author on reasonable request.

| **Characteristic** | **All patients** | *CDKN2A HETLOSS* | *CDKN2A HOMDEL* | *CDKN2A intact* |
|---|---|---|---|---|
| n | 69 (100%) | 25 (36%) | 17 (25%) | 27 (39%) |
| age at surgery (years) | 62 (±15) | 62(±18) | 60(±10) | 66(±14) |
| Sex | | | | |
| *Male* | 44 (64%) | 19 (76%) | 9 (53%) | 16 (59%) |
| *Female* | 25 (36%) | 6 (24%) | 8 (47%) | 11 (41%) |
| Ethnicity | | | | |

|  |  |  |  |  |  |
|---|---|---|---|---|---|
|  | *Asian* | 1 (1%) | 0 (0%) | 0 (0%) | 1 (4%) |
|  | *Black* | 4 (6%) | 2 (8%) | 1 (6%) | 1 (4%) |
|  | *Hispanic* | 2 (3%) | 1 (4%) | 1 (6%) | 0 (0%) |
|  | *Other* | 1 (1%) | 0 (0%) | 0 (0%) | 1 (4%) |
|  | *White* | 61 (88%) | 22 (88%) | 15 (88%) | 24 (88%) |
| Genetic profile |  |  |  |  |  |
|  | *CDKN2A HETLOSS* | 25 (36%) |  |  |  |
|  | *CDKN2A HOMDEL* | 17 (25%) |  |  |  |
|  | *CDKN2A intact* | 27 (39%) |  |  |  |
|  | *EGFR amplified* | 42 (61%) | 9 (36%) | 11 (65%) | 6 (22%) |
|  | *EGFR not amplified* | 26 (38%) | 16 (64%) | 5 (29%) | 21 (78%) |
|  | *EGFR unknown* | 1 (1%) | 0 (0%) | 1 (6%) | 0 (0%) |

**Table. 1)** Description of patient characteristics

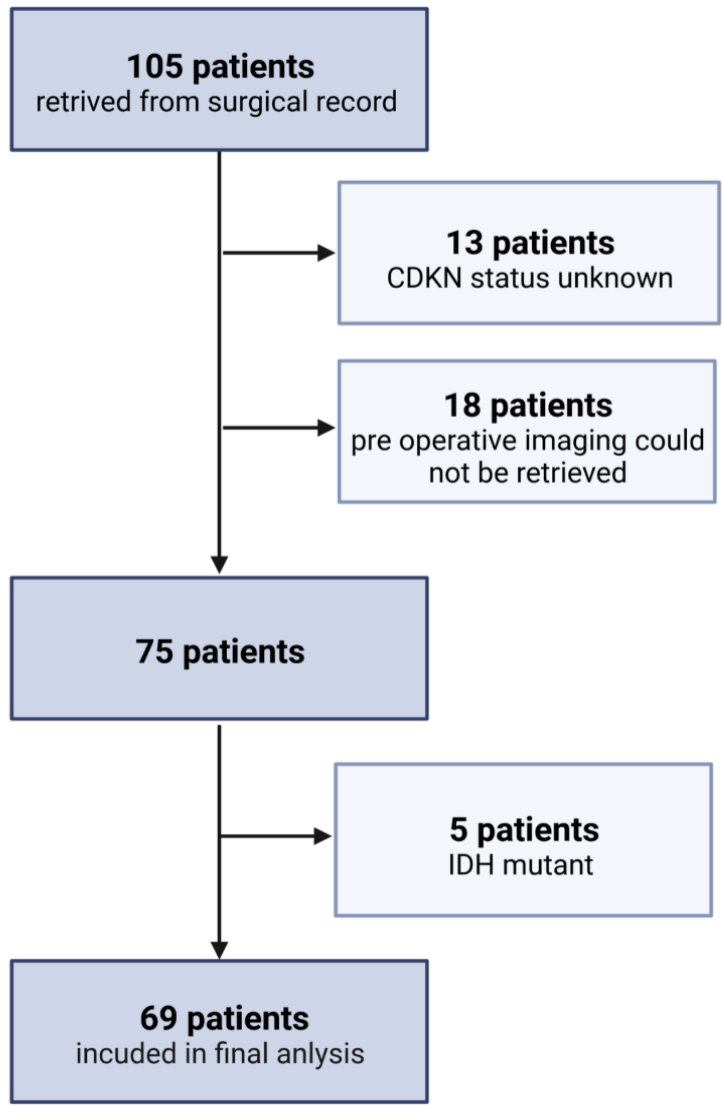

**Fig. 1)** Flowchart of patients in our analysis.

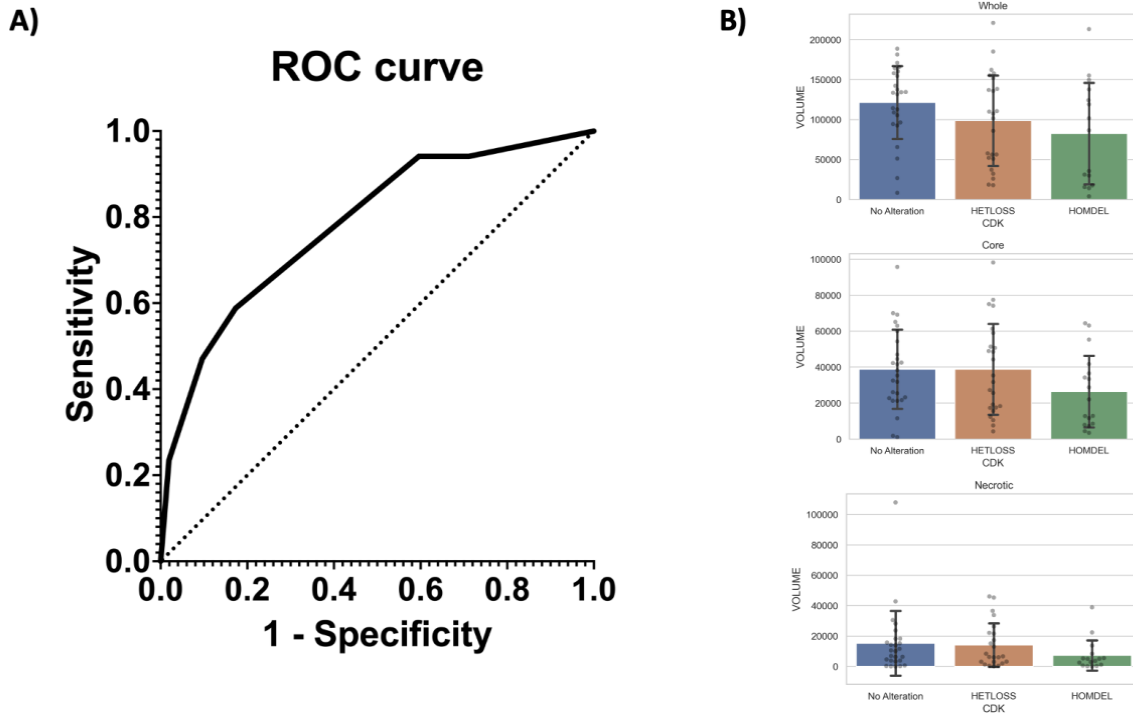

**Fig. 2) A)** Logistic regression model for prediction of homozygous deletion (HOMDEL) of CDKN2A. In a logistic regression model pial invasion was predictive for HOMDEL (Area under the ROC curve: 0,7822 ± 0.07) as tumors with pial invasion were eight times more likely to be HOMDEL, even after adjusting for deep white matter and subependymal invasion as relevant covariables (OR: 8.1, 95% CI: 1.8-53.2; p<0.012) **B)** Mean values of Whole-, Core- and Necrotic volumes in cubic millimeters based on automated segmentation, differentiated by no alteration, homozygous deletion (HOMDEL) and heterozygous loss (HETLOSS) in CDKN2A.

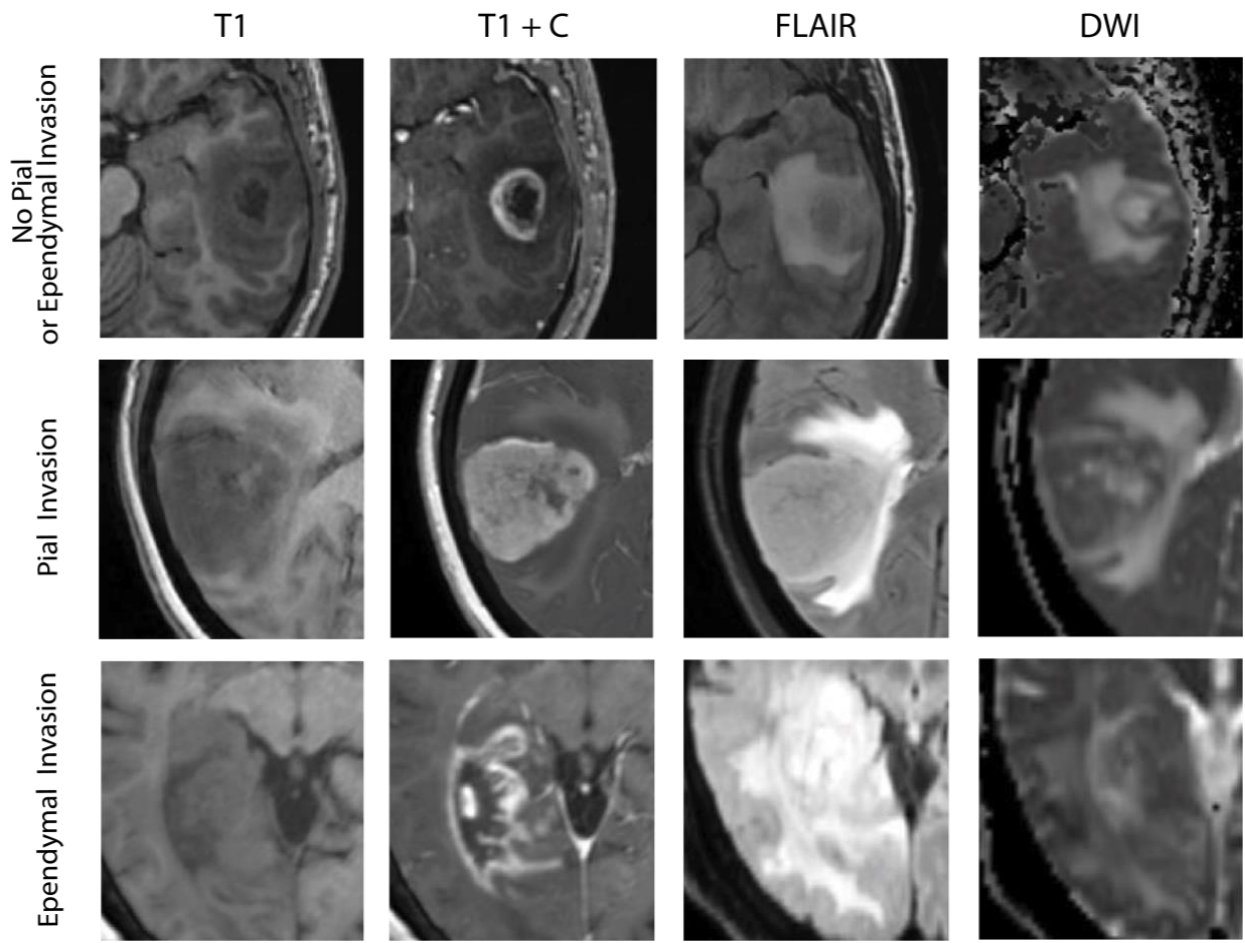

**Fig. 3)** Visualization of MRI shows no pial or subependymal invasion, pial invasion, and subependymal invasion.

FLAIR = Fluid attenuated inversion recovery, DWI = diffusion-weighted imaging

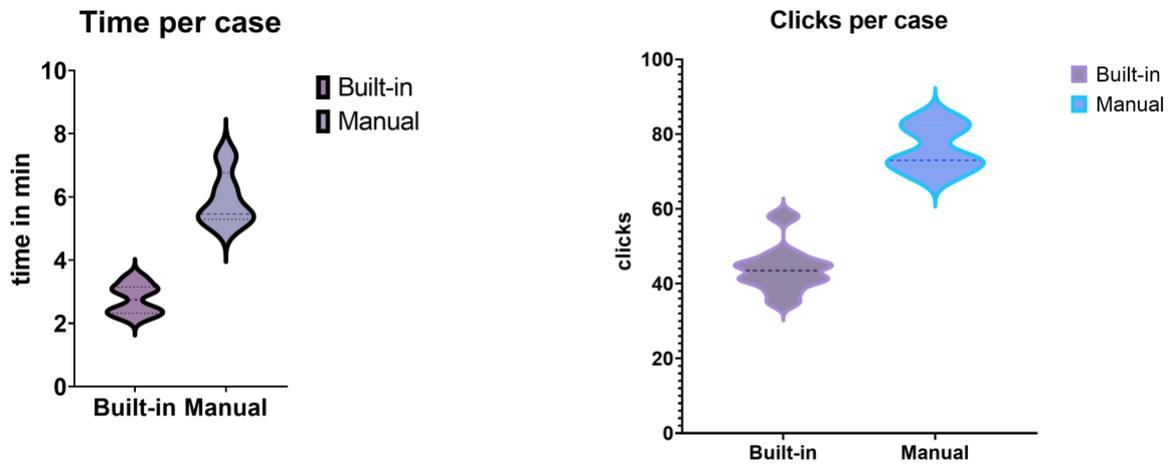

**Fig. 4)** Comparison of manual VASARI scoring with built-in forms within PACS. Shown are the median and respective quartiles. The difference between built-in and manual measurements was statistically significant regarding the time per case (p<0.0001) and clicks per case (p<0.0001). This highlights the workflow inefficiencies of manual assessment of VASARI forms using separate scoring modalities compared to native, built-in FHIR forms within PACS.